\documentclass[12pt]{article}
\usepackage[utf8]{inputenc}
\usepackage{graphicx}
\usepackage{natbib}
\usepackage{color}
\usepackage{amsmath}
\usepackage{amsfonts}
\usepackage{hyperref}
\usepackage[algo2e]{algorithm2e} 
\usepackage{algorithm,algpseudocode}
\usepackage{verbatim}
\usepackage{caption}
\usepackage{setspace}
\usepackage{subcaption}
\usepackage{multirow}
\usepackage[left=1in,right=1in,top=1in,bottom=1in]{geometry}
\newcommand{\bc}{\boldsymbol c}

\newcommand{\bA}{{\bf A}}

\newcommand{\bE}{{\bf E}}

\title{Core-periphery identification in massive networks}
\author{Eric Yanchenko \\ Akita International University\\
eyanchenko@aiu.ac.jp}

\begin{document}

\maketitle
\thispagestyle{empty}


\begin{abstract}
\noindent
Modern networks can be huge with millions or even billions of nodes and edges. Thus, algorithms must be capable of scaling to such large networks in order to be practically useful. In this work, we are interested in developing an algorithm to identify core-periphery structure in massive networks. Core-periphery structure is a meso-scale feature where nodes are grouped into a densely connected core or sparsely connected periphery. To identify such structures in large networks, we propose a divide-and-conquer algorithm. The key feature of our algorithm is leveraging the edge list representation of the network, instead of the adjacency matrix, as it tends to be faster and makes a more efficient use of memory. We apply the proposed algorithm to synthetic and real-world data, notably  demonstrating its performance on a real-world network with almost 14 million edges without loading the entire network into memory. 

\end{abstract}

\section{Introduction}\label{sec:intro}

With the modern proliferation of data, networks today can be massive. Across numerous domains, researchers now have access to huge networks which require scalable algorithms and tools. In particular, practitioners are often interested in identifying meso-scale structures in networks such as community structure \citep{newman2004finding, Fortunato:2010aa} or core-periphery (CP) structure \citep{BORGATTI2000, yanchenko2023core}. We focus on CP structure where networks have two groups a nodes: a densely connected core and a more sparsely connected periphery that forms edges with core nodes more frequently than with other periphery nodes. Analyzing this structure allows for, e.g., identification of the most influential nodes, and has been studied in a range of applications including economic \citep{krugman1996self, magone2016core}, social \citep{cattani2008, yang2018structural}, research citations \citep{sedita2020invisible, wedell2022center}, and more. In this work, we are particularly interested in identifying CP structure in large networks.

One standard approach to handle large data sets (not only networks) is sub-sampling which allows data to be analyzed in smaller pieces before combining the results for the entire data set. Sub-sampling has some unique challenges in networks due to the inherent dependence in the data as well as their complicated topologies, but previous approaches exist for community and core-periphery detection \citep{mukherjee2021two, yanchenko2022divide}. In particular, \cite{yanchenko2022divide} develops a divide-and-conquer algorithm to identify CP structure with subsequent work focusing on the effect of the sub-sampling algorithm \citep{yanchenko2025graph}. In general, it was shown that sub-sampling routines that are biased towards sampling central or high degree nodes tended to perform the best.

This approach demonstrated good empirical performance but has limited scalability to truly massive networks. The crux of the problem is that these methods use an {\it adjacency matrix} representation of the network in order to carry out the analysis. An adjacency matrix $\bA$ is an $n\times n$ matrix where $n$ is the number of nodes and $A_{ij}=1$ if nodes $i$ and $j$ have an edge, and 0 otherwise. Alternatively, many large networks are represented via an {\it edge list} where each row records the incident and source nodes connected by an edge. While an adjacency matrix approach may be sufficient for some networks, it cannot scale to extremely large networks; adjacency matrices require $O(n^2)$ storage space which can be inefficient if the network is sparse as most entries will be zero. Moreover, the algorithms mentioned above require the entire adjacency matrix to be loaded into memory in order to identify the labels, meaning that this algorithm is also limited by the memory of the computing environment.

To overcome these challenges, we propose a sub-sampling algorithm to identify CP structure in massive networks using an edge list representation. We first describe optimizing the CP metric from \cite{BORGATTI2000} using the edge list and a greedy algorithm. We show that if the network is sparse then this algorithm is faster than the analog using the adjacency matrix. To apply this algorithm to massive networks, however, we propose a divide-and-conquer algorithm by randomly sampling edges from the edge list and then finding the optimal CP labels on these sub-graphs. This process is repeated many times and the results are combined via averaging to yield a continuous measure of ``coreness'' for each node, which can be converted to binary CP labels. Our approach outperforms a similar adjacency matrix-based algorithm on synthetic networks. Moreover,  we apply our algorithm to a real-world network with almost 14 million edges without loading the entire network into the memory. Thus, our proposed approach is scalable to truly massive networks.

The rest of the paper is organized as follows. In Section \ref{sec:alg} we describe the CP objective function and greedy algorithm to optimize this metric which is then leveraged in a divide-and-conquer algorithm. The proposed algorithm is applied to simulated and real-world networks in Section \ref{sec:data}, and concluding thoughts are shared in Section \ref{sec:conc}.

\section{Methodology}\label{sec:alg}

\subsection{Notation}
Let $G$ be a graph with $n$ nodes and $m$ edges. We define $\bA\in\{0,1\}^{n\times n}$ as the adjacency matrix corresponding to graph $G$ such that $A_{ij}=1$ if nodes $i$ and $j$ are connected by and edge, and 0 otherwise.\footnote{In this work, we only consider undirected, unweighted networks without self-loops, but the ideas can easily be generalized.} Additionally, define $\bE\in\{1,\dots,n\}^{m\times 2}$ as the edge list representation of the network where each row records the nodes involved in the edge, and $E_i(j)$ corresponds to the $j$th node involved in the $i$th edge for $j\in\{1,2\}$ and $i\in\{1,\dots,m\}$. Let $\bc\in\{0,1\}^n$ be a CP assignment vector where $c_i=1$ if node $i$ is assigned to the core, and 0 otherwise. Let $k=\sum_{i=1}^n c_i$ be the number of nodes in the core, and $\alpha=k/n$ be the proportion of nodes in the core.

\subsection{Core-periphery metric}
We quantify the strength of the CP structure in the observed network using the objective function from \cite{BORGATTI2000}. For a given $\bc$, we define
\begin{equation}\label{eq:obj}
    T(\bA,\bc)
    =\frac{\sum_{i<j}(A_{ij}-\bar p)\Delta_{ij}}{\frac12n(n-1)\{\bar p(1-\bar p)\bar\Delta(1-\bar\Delta)\}^{1/2}}
\end{equation}
where $\bar p=\sum_{i<j}A_{ij}/\{n(n-1)/2\}$ is the average edge probability and $\bar\Delta=\{\frac12k(k-1)+k(n-k)\}/\{\frac12n(n-1)\}$ where $\Delta\in\{0,1\}^{n\times n}$ corresponds to the ``ideal'' CP structure defined as $\Delta_{ij}=c_i+c_j-c_ic_j$. Thus, \eqref{eq:obj} quantifies how closely the observed network matches one with a perfect CP structure. While we have defined this metric in terms of the adjacency matrix, $\bA$, we describe below how to calculate it using the edge list, $\bE$, as well. The CP vector $\bc$ is often unknown in practice so we treat \eqref{eq:obj} as an objective function to find the optimal CP labels, i.e., 
\begin{equation}\label{eq:opt}
    \hat\bc=\arg\max_{\bc}T(\bA,\bc).    
\end{equation}
The majority of this work is focused on finding the solution in \eqref{eq:opt}. In particular, this is a combinatorial optimization problem over a space of $2^n$ possible solutions. For even modest $n$, e.g., $n=100$, the entire search space is enormous so a brute-force algorithm to find the global maximum is impossible. Instead, we seek to employ heuristics to approximate the solution. 

\subsection{Objective function evaluation}

Any algorithm which seeks to find the optimal solution in \eqref{eq:opt} must first evaluate the objective function in \eqref{eq:obj}. We begin by reviewing how to do this using the graph's adjacency matrix \citep{yanchenko2026label} before turning to the edge list. Given an adjacency matrix $\bA$ and CP labels $\bc$, the key step needed to evaluate \eqref{eq:obj} involves computing the sum in the numerator, which can be re-written as
\begin{equation}\label{eq:obj2}
    T(\bA,\bc)
    =\frac{\sum_{i<j}A_{ij}\Delta_{ij}-\{\frac12k(k-1)+k(n-k)\}\bar p\bar \Delta}{\frac12n(n-1)\{\bar p(1-\bar p)\bar\Delta(1-\bar\Delta)\}^{1/2}}.
\end{equation}
Clearly, the time consuming step is $\sum_{i<j}A_{ij}\Delta_{ij}$ which requires looping over both rows and columns of the adjacency matrix, i.e., $O(n^2)$ flops. Please see the Appendix for full details.

Given a current value of the objective function, \cite{yanchenko2026label} describe an efficient update to the value for new labels in $O(n)$ operations, instead of a naive implementation requiring $O(n^2)$. Specifically, let $\bc$ be the current labels and $M:=M(\bc)$ be known where we emphasize the dependence of $M(\bc)$ on $\bc$. Furthermore, let $\bc'$ be labels defined as $c'_i=1-c_i$ for $i\in\{1,\dots,n\}$ and $c'_j=c_j$ for all $j\neq i$. In words, $\bc'$ differs from $\bc$ at only a single node $i$. The key observation from \cite{yanchenko2026label} is that $M(\bc')$ can be efficiently updated using the value of $M(\bc)$. Specifically,

\begin{align}\label{eq:comp}\notag
    M(\bc')
    &=\sum_{j=1}^n\sum_{k=1}^{j-1} A_{jk}\Delta_{jk}(\bc') \\ \notag 
    &=\sum_{j\neq i}^n\sum_{k=1}^{j-1}A_{jk}\Delta_{ik}(\bc')
    +\sum_{k=1}^{i-1} A_{ik}\Delta_{ik}(\bc')\\ \notag
    &=\sum_{j\neq i}^n\sum_{k=1}^{j-1}A_{jk}\Delta_{jk}(\bc) 
    +\sum_{k=1}^{i-1} A_{ik}\Delta_{ik}(\bc')\\ \notag
    &= \left(M(\bc) -\sum_{k=1}^{i-1} A_{ik}\Delta_{ik}(\bc)\right) +\sum_{k=1}^{i-1} A_{ik}\Delta_{ik}(\bc')\\\notag
    &=M(\bc)+\sum_{k=1}^{i-1}A_{ik}\{\Delta_{ik}(\bc')-\Delta_{ik}(\bc)\}\\
    &=M(\bc)+\sum_{k=1}^{i-1}A_{ik}(c_i'-c_i)(1-c_i).
\end{align}

\noindent
Since we know $M(\bc)$, we simply need to evaluate the final sum in \eqref{eq:comp} which only requires $O(n)$ flops. Thus, given $T(\bA,\bc)$ (and $M(\bc)$), we can find $T(\bA,\bc')$ in $O(n)$ operations. These steps are again laid out in the Appendix.

Evaluating the Borgatti and Everett metric is similar when using an edge list representation of $G$. We can re-write \eqref{eq:obj} as
\begin{equation}\label{eq:obj3}
    T(\bA,\bc)
    =\frac{\sum_{i=1}^m M_i-\{\frac12k(k-1)+k(n-k)\}\bar p\bar \Delta}{\frac12n(n-1)\{\bar p(1-\bar p)\bar\Delta(1-\bar\Delta)\}^{1/2}},
\end{equation}
where $M_i=1$ if $c_{E_i(1)}+c_{E_i(2)}>1$ (least one node assigned to the core), and 0 otherwise. The sum in the numerator can be found by looping through the edge list and checking if either incident node is assigned to the core in $O(m)$ flops. These steps are detailed in Algorithm \ref{alg:edge}.

\begin{algorithm} 
\caption{Objective function evaluation (edge list)}
\label{alg:edge}
\begin{algorithmic}
\Require{$E$ edge list, $c$ core-periphery vector.} 
\Ensure{$T$ objective function value}
\Statex
\Function{objFunEdge}{$E, c$}
\State {$n \gets \text{length}(c)$}
\State {$k \gets \sum_{i=1}^n c(i)$}
\State {$m \gets \text{numRows}(E)$}
\State {$\bar\Delta \gets (k(k-1)/2+k(n-k))/(n(n-1)/2)$}
\State {$\bar p \gets m/(n(n-1)/2)$}

\State{Initialize $M\gets 0$}

\State{\For{$i$ in $1:m$}{

    \If{$c(E(i,1))+c(E(i,2))>0$}{
        $M\gets M+1$\;
    }

}
}

\State{$T\gets \frac{M-\{\frac12k(k-1)+k(n-k)\}\bar p\bar \Delta}{\frac12n(n-1)\{\bar p(1-\bar p)\bar\Delta(1-\bar\Delta)\}^{1/2}}$}

\State \Return {$T$}
\EndFunction
\end{algorithmic}
\end{algorithm}

\subsection{Greedy algorithm}
Now that we can evaluate the objective function using either the adjacency matrix or edge list, we turn our attention to finding the optimal solution in \eqref{eq:opt}. To do so, we employ the greedy, label switching algorithm from \cite{yanchenko2026label}. Given a graph $G$, we randomly initialize the CP labels $\bc$ and evaluate $T(\bA,\bc)$. Then for each node $i$, we swap its assignment, i.e., from core to periphery or periphery to core, and define this new, proposed label as $\bc'$. Clearly, $c_j'=c_j$ for all $j\neq i$ and $c_i'=1-c_i$. The objective function is then evaluated on these new labels $T(\bA,\bc')$, and the proposed swap is kept if $T(\bA,\bc')>T(\bA,\bc)$. The steps are similar for the edge list representation, with full details in Algorithm \ref{alg:greedyE}. The adjacency matrix-based algorithm is left to the Appendix.

For the adjacency matrix representation, initially evaluating the objective function requires $O(n^2)$ flops, but updating its value in the loop only takes $O(n)$ flops. This is performed for each node so a single pass of the algorithm requires $O(n^2)$ operations. For the edge list algorithm, initially evaluating the objective function takes $O(m)$ operations. Since the objective function is calculated from scratch each time a node is swapped, this also takes $O(m)$ flops such that a single pass takes $O(nm)$ operations. Ideally, we could also derive a clever update of the objective function for the edge list representation as in \eqref{eq:comp}, but we are unaware how to do this at this point. 

The advantages of the edge list algorithm compared to that of the adjacency matrix begin to become apparent from this discussion. First, the relationship between $m$ and $n$ governs the relative speed of the algorithms. If the number of nodes is significantly larger than the number of edges, i.e., $m=o(n)$, then the edge list approach will be faster. In many theoretical proofs for networks \citep[e.g.,][]{bickel2009nonparametric}, the degree of the nodes is assumed to be $O(1)$ such that number of edges in the network grows linearly with $n$, i.e., $m=O(n)$. In this case, the algorithms should be approximately equivalent in terms of speed, but the edge list approach still maintains some advantages, e.g., in terms of storage, the adjacency matrix is $O(n^2)$ while the edge list is only $O(m)$ and, in virtually all real-world networks, $m=o(n^2)$ such that the edge list makes a much more efficient use of space.

\begin{algorithm}
\caption{Greedy algorithm (edge list)}
\label{alg:greedyE}
\begin{algorithmic}
\Require{$E$ edge list} 
\Ensure{$c$ core-periphery labels}
\Statex
\Function{greedyEdge}{$E$}
\State {Randomly initialize labels $c$}
\State{$T\gets${\sc objFunEdge}$(E,c)$}
\State { $run \gets 1$}
\State{\While{$run>0$}{
 Set $run\gets0$\; 
 
 Randomly order nodes\;
 
  \For{$i$ in $1:n$}{

  $c'\gets c$; $c'_i\gets 1-c_i$\;

  $T'\gets$\sc{objFunEdge}$(E,c')$
  
  \If{$T' > T$}{
  
  $c\gets c'$; $T\gets T'$\;

  $run \gets 1$\;
 }
   
  }
}
}

\State \Return {$c$}
\EndFunction
\end{algorithmic}
\end{algorithm}

\subsection{Divide-and-conquer algorithm}
While the greedy algorithm is fast for medium and large networks, it is still too slow for extremely large networks since the edge list algorithm scales linearly with the number of nodes times the number of edges. Moreover, this algorithm ostensibly requires the entire network to be loaded into memory, so it cannot be implemented when the network is so large that this is not possible. These two hurdles motivate the need for a divide-and-conquer approach for large networks, which we adapt from \cite{yanchenko2022divide, yanchenko2025graph}. Specifically, the user chooses $B\in\mathbb Z^+$, the number of sub-samples and $q\in(0,1)$, the proportion of the total edges to be sampled. Then a sub-graph with $qm$ edges is randomly sampled and the optimal CP labels are calculated. This is repeated $B$ times and the proportion of times that each node was assigned to the core out of all sub-samples is reported. Full details can be found in Algorithm \ref{alg:dac}.

\begin{algorithm}
\SetAlgoLined
\KwResult{Core-periphery proportions $\hat{c}$}
 {\bf Input: }Edge list $E$, proportion of edges to sub-sample $q\in(0,1)$, number of sub-samples $B\in\mathbb Z^+$\;
 
  \For{$B$ times}{
  Randomly sample $qm$ edges in $E$ to obtain sub-graph $E^{(b)}\in\{1,\dots,n\}^{qm\times 2}$\;

  $\hat{\bc}^{(b)}\gets${\sc greedyEdge}$(E^{(b)})$\;}
  
  $\hat c_i=\displaystyle\frac{1}{B}\sum_{b=1}^B\hat c_i^{(b)}$\;
 
\caption{Divide-and-conquer for core-periphery structure}
\label{alg:dac}
\end{algorithm}

\subsection{Discussion}\label{sec:disc}

The divide-and-conquer Algorithm \ref{alg:dac} leveraging the edge list provides various advantages over the base algorithm (Algorithm \ref{alg:greedyE}) as well as the adjacency matrix-based divide-and-conquer algorithms of \cite{yanchenko2022divide, yanchenko2025graph}. First, it is more computationally efficient than the base algorithm. Given a graph with $n$ nodes and $m$ edges, Algorithm \ref{alg:greedyE} requires $O(nm)$ operations for a single pass. In the divide-and-conquer algorithm, we sample $qm$ edges to construct a sub-graph which means that the function {\sc objFunEdge}$()$ (Algorithm \ref{alg:edge}) requires $O(qm)$ operations to be evaluated. For a random sub-graph, we do not know how many nodes, $n_s$, will be sub-sampled, but it is upper-bounded by $2qm$. Thus, for a single sub-graph, $n_s=O(qm)$ such that a single pass of Algorithm \ref{alg:greedyE} requires $O((qm)^2)$ operations. Sub-sampling is carried out $B$ times, resulting in $O((qm)^2B)$ operations. The divide-and-conquer algorithm, however, is embarrassingly parallel so given $d$ cores, the total complexity is
$$
    O\left(\frac{(qm)^2B}{d}\right).
$$
Thus, $R$, the relative speed of Algorithm \ref{alg:dac} to Algorithm \ref{alg:greedyE}, is 
$$
    R
    =\frac{(qm)^2B/d}{nm},
$$
which depends on the choices of $q,B$ and $d$, and where $R<1$ means the divide-and-conquer algorithm is faster than the base algorithm. Below, we suggest setting $q$ and $B$ such that $qB=O(1)$. Additionally, large graphs are often such that $m=O(n)$ so $R$ simplifies to
$$
    R=\frac{q}{d}.
$$
If, e.g., $q=10^{-3}$ and $d=10$, then $R=10^{-4}$ such that the divide-and-conquer algorithm is expected to be $10^{4}$ times faster than the base algorithm.

While superficially similar to the adjacency matrix-based divide-and-conquer algorithms of \cite{yanchenko2022divide, yanchenko2025graph}, the edge list approach of Algorithm \ref{alg:dac} possesses some unique advantages. The first comes from an efficient use of data storage. To see this, consider the following example. Recall that many large networks are sparse, where the number of edges is of the same order as the number of nodes. This means that to ensure (on the order of) 100 edges in a sub-graph, one would need to sample (on the order of) 100 nodes (assuming random node sampling), leading to a $100\times100$ adjacency matrix. Clearly, this requires $100\times 100=10^4$ bytes of storage (assuming 8 bit numbers), while the edge list approach only requires $2\times 100=200$ bytes. Thus, in general, the adjacency matrix divide-and-conquer algorithm will require much larger sub-graphs to obtain comparable performance, meaning it will be slower than the edge list approach. While this can be somewhat mitigated with alternative sampling approaches, we show in the simulation study that this problem still persists. Moreover, the edge list approach of Algorithm \ref{alg:dac} can be applied to truly massive networks which cannot even be loaded into the memory at a single time. Indeed, edges can be sampled directly from the edge list data structure without loading the entire list into memory, which we demonstrate in Section \ref{sec:data_real}. Conversely, the adjacency matrix approach as applied in \cite{yanchenko2022divide, yanchenko2025graph} requires the entire adjacency matrix to be loaded, greatly limiting its scalability.

In \cite{yanchenko2025graph}, various sub-sampling algorithms are considered to construct the sub-graphs. Algorithm \ref{alg:dac} clearly uses random edge sampling which is known to have a bias of sampling nodes with larger degrees \citep{ribeiro2010estimating}, but this is actually an advantage for our method since higher degree nodes are more likely be core nodes. Indeed, \cite{ribeiro2010estimating} show that random edge samplers better estimate the tails of the degree distribution which likely corresponds to core nodes in CP structure. While the random edge sampler was shown to perform well in \cite{yanchenko2025graph}, we stress that these two approaches have an important distinction. In the adjacency matrix approach, we randomly sample an edge, keep the two incident nodes and then include {\it all other edges} for those nodes. This is sometimes called the {\it graph induction} step \citep{ahmed2011network}. In Algorithm \ref{alg:dac}, however, we randomly sample edges but {\it do not} include all other edges incident to those nodes. In other words, we do random edge sampling {\it without} induction. While induction may be preferable, it is not obvious how to do this with an edge list representation without scanning over the entire edge list, something that is not feasible if, e.g., the edge list cannot be loaded into memory.

One concern about not using the induction step is that it is highly possible that there will only be a single edge for each sampled node in the sub-graph such that every node has degree one. If this is the case, however, then there is clearly no meaningful CP structure in the sub-graph. To mitigate this problem, we must ensure that $q$ is large enough such that core nodes sampled in the sub-graph have a sufficient number of edges to be identified as such. With this in mind, we propose the following heuristic to set $q$. Let $q$ vary in some pre-specified range, e.g., $\log_{10}q\in\{-3,-4,\dots,-7\}$. Then for each value of $q$, randomly sample $qm$ edges from the graph and record if at least one node has multiple edges in the sub-graph. Repeat this process $N$ times and, for each value of $q$, compute the proportion of sub-samples for which the sub-graphs contained one or more nodes with multiple edges. Then select $q$ as the smallest value such that this proportion is at least 90\%. This ensures that a minimum of 90\% of the sub-samples will contain at least one node with multiple edges. We employ this procedure in the real-data section with $N=100$. Finally, to set $B$, note that $qB$ is the expected number of times that an edge is sampled, so after finding $q$, we suggest setting $B=q^{-1}$ such that each edge is expected to be sampled once. Given greater computational resources, $B$ can be larger but we suggest it is no smaller than $q^{-1}$.

\section{Experiments}\label{sec:data}

\subsection{Synthetic data}
In this section, we study the performance of the proposed algorithm on synthetic and real-world networks. All experiments were carried out on 2024 Mac Mini computers with Apple M4 chip and 16 GB of memory. For the divide-and-conquer algorithms, we parallelized the code across 9 cores. Code was written in R and is available at the author's GitHub: \url{https://github.com/eyanchenko/CPDACedge}. All results were averaged over 100 Monte Carlo replications. 

\begin{figure}
    \centering
    \includegraphics[width=0.5\linewidth]{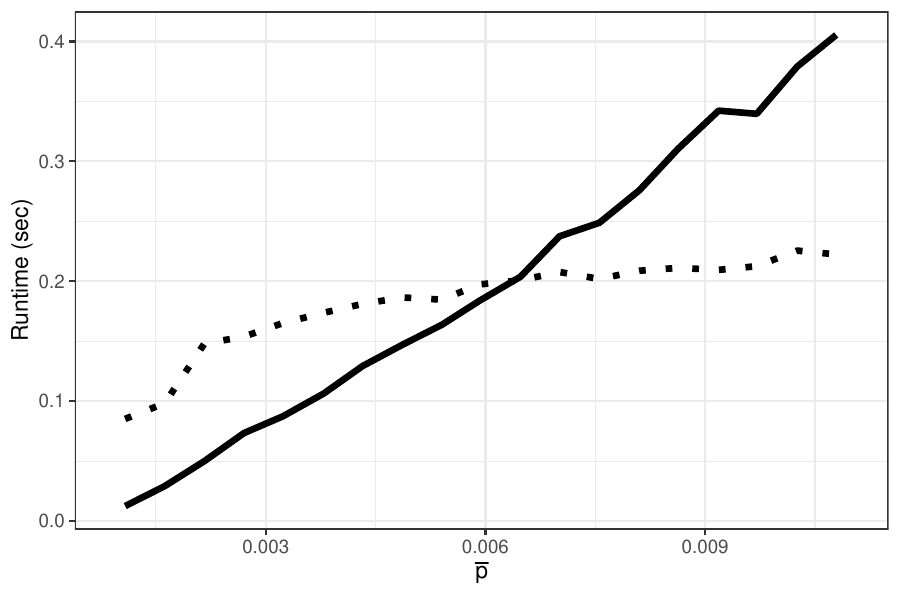}
    \caption{Comparing runtime against density for the edge list-base greedy algorithm (solid line) and adjacency matrix-based greedy algorithm of \cite{yanchenko2026label} (dotted line).}
    \label{fig:sims_sparsity}
\end{figure}

First, we compare Algorithm \ref{alg:greedyE}, which uses the edge list of the network, with the corresponding algorithm based on the adjacency matrix from, e.g., \cite{yanchenko2026label}. We generate networks from a stochastic block model (SBM) \citep{holland83} with known CP structure where the probability of a core-core, core-periphery and periphery-periphery edge is $p_{11},p_{12},p_{22}$, respectively, where $p_{11}>p_{12}>p_{22}$. We set $n=1000$, $\alpha=0.01$, $p_{11}=\xi$, $p_{12}=0.5\xi$ and $p_{22}=0.1\xi$, and vary $\xi\in\{0.010,0.015,\dots, 0.100\}$ to vary the density of the network. In Figure \ref{fig:sims_sparsity}, we plot the runtime against the average network density ($\bar p=\sum_{ij}A_{ij}/\{n(n-1)\})$ for each value of $\xi$. The computation time increases for both algorithms as the density increases with Algorithm \ref{alg:greedyE} increasing at a faster rate. For sparser networks ($\bar p\leq 0.006$), the edge list-based approach is faster. This is to be expected as Algorithm \ref{alg:greedyE} scales with the number of edges in the network which increases as $\bar p$ increases.

Next, we study the divide-and-conquer method of Algorithm \ref{alg:dac}. We compare its performance with the divide-and-conquer algorithm of \cite{yanchenko2022divide, yanchenko2025graph} (using the adjacency matrix), as well as the base greedy algorithm (Algorithm \ref{alg:greedyE}). For this experiment, we are primarily interested in studying the effects of the sub-sampling algorithm. As previously discussed, the random edge sampler of \cite{yanchenko2025graph} performs an induction step after sampling the edges to ``fill-in'' the rest of the sub-graph, a step that is not performed in Algorithm \ref{alg:dac}. For the edge list divide-and-conquer algorithm, we set $q=0.01$, and to make a fair comparison with the algorithm from \cite{yanchenko2025graph}, we select $q$ (the proportion of nodes sampled) for the adjacency matrix-base algorithm such that the number of edges in the sub-graphs is approximately equal to that of the edge list approach. 

\begin{figure}
    \centering
    \includegraphics[width=0.75\linewidth]{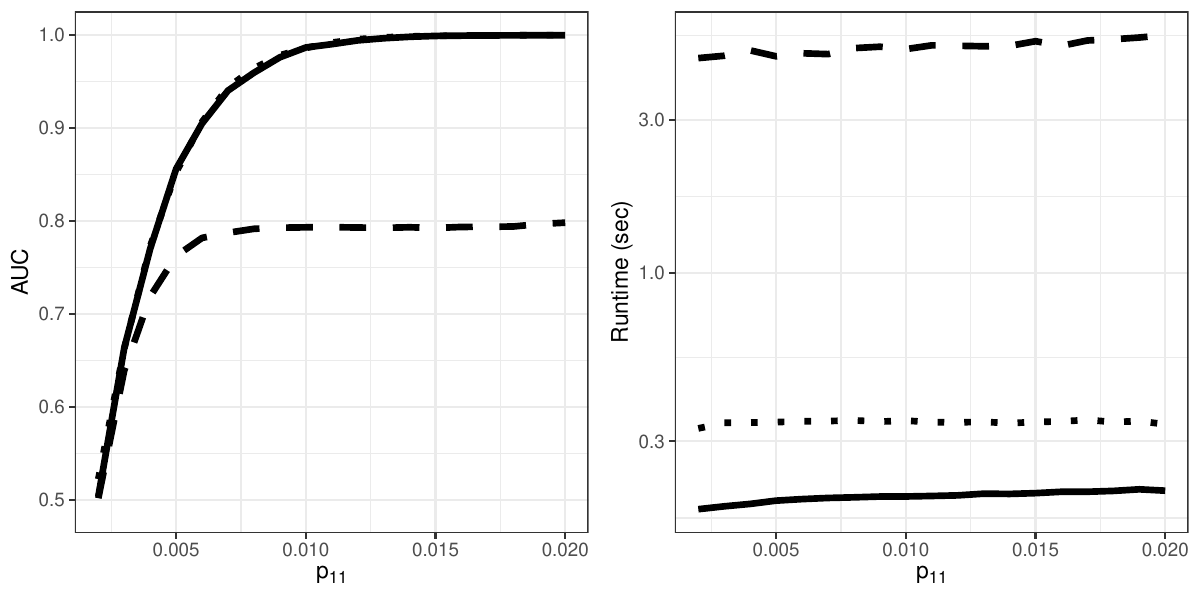}
    \caption{AUC and runtime against the strength of the core-periphery structure for the edge list-based greedy algorithm (dashed line) and divide-and-conquer algorithms using the adjacency matrix (dotted line) and edge list (solid line).}
    \label{fig:sims_rho}
\end{figure}

\begin{figure}
    \centering
    \includegraphics[width=0.75\linewidth]{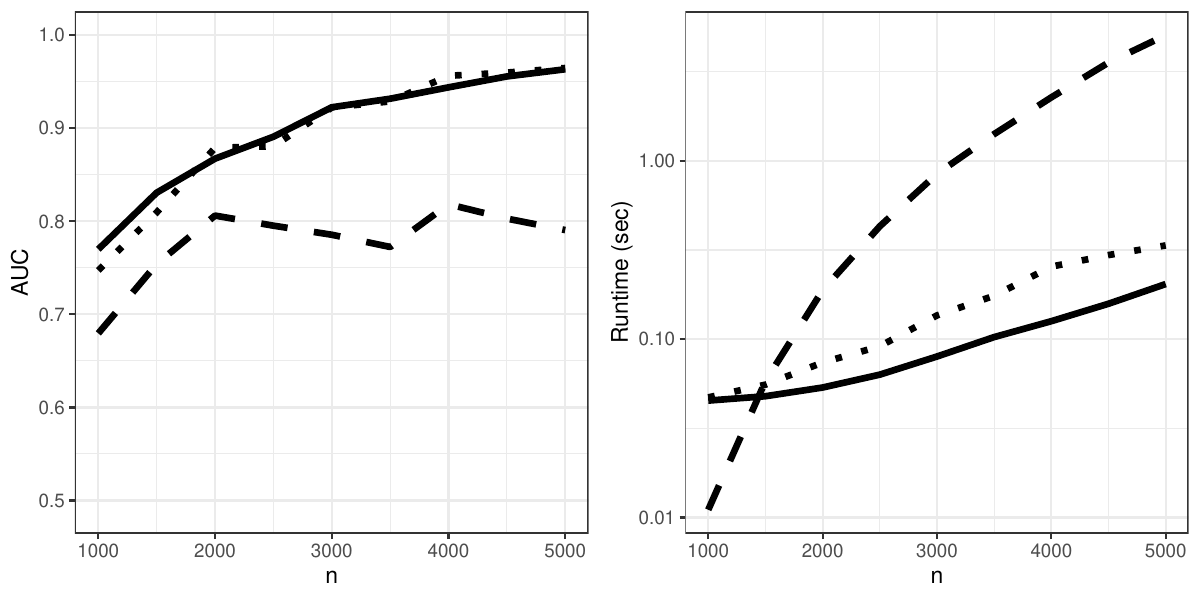}
    \caption{AUC and runtime against the number of nodes for the edge list-based greedy algorithm (dashed line) and divide-and-conquer algorithms using the adjacency matrix (dotted line) and edge list (solid line).}
    \label{fig:sims_varyn}
\end{figure}

We generate networks from an SBM with $n=5000$, $\alpha=0.01$, $p_{11}\in\{0.002, 0.003,\dots, 0.020\}$, $p_{12}=p_{11}/2$ and $p_{22}=0.001$. We set $B=100$ (number of sub-samples) and record the area under the receiver-operator curve (AUC) and run-time for increasing $p_{11}$. The results are in Figure \ref{fig:sims_rho}. Both divide-and-conquer algorithms outperform the base algorithm in terms of speed and accuracy. While the speed improvement is expected, the accuracy gains are somewhat surprising and are likely due to the sub-sampling step ``averaging out'' some of the noise in the network. Similar findings were reported in \cite{yanchenko2025graph}. While both divide-and-conquer algorithms have a comparable performance in terms of AUC (monotonic increase with $p_{11}$),  Algorithm \ref{alg:dac} is faster than the adjacency matrix-based algorithm. The results are similar when varying the size of the network. We keep the same settings as above but now we fix $p_{11}=0.008$, $p_{12}=0.004$, $p_{11}=0.001$ and vary $n\in\{1000,1500,\dots,5000\}$. The results are in Figure \ref{fig:sims_varyn}. The divide-and-conquer methods again have a similar AUC that is larger than that of the greedy algorithm, and the edge list-based algorithm is the fastest for $n\geq2000$.

\subsection{Real-world networks}\label{sec:data_real}
Finally, we demonstrate the performance of the proposed algorithm on four real-world networks. Specifically, we consider Amazon \citep{leskovec2007dynamics}, YouTube \citep{yang2012defining}, Twitch \citep{rozemberczki2021twitch} and Google \citep{leskovec2012learning} networks. Summary statistics for each are available in Table \ref{tab:real} where the largest (Google) has almost 14 million edges. All networks are available on the SNAP network repository \citep{snapnets} and were pre-processed to remove self-loops, edge directions, etc.

For each network, we apply Algorithm \ref{alg:dac} to obtain the CP proportions for each node. We follow the procedure described in Section \ref{sec:disc} to choose $q$. 
Since ground-truth CP labels are unknown, we convert these continuous proportions to binary labels and compute the objective function value in \eqref{eq:obj}. To do so, we order nodes based on their proportions in descending order. Then one at a time, we add the node with the largest proportion to the core and evaluate the objective function. We then add the node with next largest proportion to the core and again calculate the objective function. We repeat this process and the labels which correspond to the largest objective function value are kept as the optimal labels. As a baseline comparison, we also consider a degree ranking approach where we rank nodes by their degree and then carry out the same procedure as described above. In this sense, both methods yield a ranking of a nodes based on ``coreness'' and we compare to see which method yields a larger value of the objective function.

The results are in Table \ref{tab:real}. We report the objective function value, number of nodes assigned to the core and computing time for each method. For each graph, the proposed algorithm yields significantly larger objective function values, typically at least one order of magnitude larger. For example, on the Twitch network, Algorithm \ref{alg:dac} yields an objective function value more than 20 times greater than that of the degree-ranking heuristic. We note that some of the raw values of the objective function are small but this is not necessarily a flaw in the algorithm as it likely means that this network does not possess a strong CP structure. Regardless, Algorithm \ref{alg:dac} is able to find labels which leader to much larger objective function values than simple degree ranking.

Finally, we highlight that for the Google network, the edge list took up about 0.15 GB of memory. While this can easily be loaded into R, we sought to demonstrate that the proposed method does not require that the entire edge list be loaded at once. Thus, we used the LaF \citep{laf} package in R to sample edges directly from the edge list .txt file without loading it into memory. Thus, we consider this as a proof-of-concept showing that even using a personal desktop machine, our proposed algorithm is able to handle massive graphs.

\begin{table}[]
    \centering
    \begin{tabular}{lcc|ccccc}
        Network  & $n$ & $m$ & Method & $\log_{10}q$ & Obj & $k$ & Time\\\hline  
        Amazon & 0.4M & 2.3M &&&\\
        &&&DAC & $-3$ & 0.006 & 1132 & 3.4\\
        &&&Deg &  & 0.001 & 42 & $<0.01$\\  
        YouTube & 1.1M & 3.0M &&&\\ 
        &&&DAC & $-4$ & 0.018  & 7 & 2.2 \\
        &&&Deg &   & 0.002 & 1356 & $<0.01$\\  
        Twitch  & 1.7M & 6.8M &&&\\
        &&&DAC & $-4$ & 0.083 & 177 & 1.5\\
        &&&Deg &      & 0.004 & 338 & $<0.01$\\ 
        Google & 1.1M & 14M &&&\\ 
        &&&DAC & $-5$ & 0.116  & 1712 & 2.2 \\
        &&&Deg &   & 0.046 & 3405 & $<0.01$
    \end{tabular}
    \caption{Real-data results. Network: network name. $n$: number of nodes (in millions). $m$: number of edges (in millions). Method: DAC corresponds to the edge list divide-and-conquer algorithm, and Deg represents the degree ranking method. Obj: objective function value calculated in \eqref{eq:obj}. $k$: number of nodes assigned to the core. Time: computing time in hours.}
    \label{tab:real}
\end{table}

\section{Conclusion}\label{sec:conc}
In this work, we proposed a divide-and-conquer algorithm to identify CP structure in massive networks using the edge list representation of the network. This approach allows for computational speed ups as well as more efficient use of memory compared to an adjacency matrix representation. Indeed, on synthetic and real-world networks, the proposed algorithm performed well and was applied to a network with almost 14 million edges without needing to load the entire network into memory. At this point, the only limitations on applying the algorithm to even bigger networks are the number of computing cores available as well as the speed with which the edge list can be queried. Indeed, for even larger networks with edge lists on the order of gigabytes, fast sub-sampling of the edge list is necessary to scale the algorithm. Other future work could look into a more theoretically guaranteed approach to choose the hyper-parameters in the algorithm, specifically the size of the sub-graph.

\bibliographystyle{apalike}
\bibliography{refs}

\clearpage

\section*{Appendix}

\begin{algorithm} 
\caption{Objective function evaluation (adjacency matrix)}
\label{alg:adj}
\begin{algorithmic}
\Require{$A$ adjacency matrix, $c$ core-periphery vector.} 
\Ensure{$T$ objective function value, $M$ number of core-core and core-periphery edges}
\Statex
\Function{objFunAdj}{$A, c$}
\State {$n \gets \text{length}(c)$}
\State {$k \gets \sum_{i=1}^n c(i)$}
\State {$\bar\Delta \gets (k(k-1)/2+k(n-k))/(n(n-1)/2)$}

\State {Initialize $M\gets 0;\ \bar p\gets 0$}

\State{\For{$i$ in $2:n$}{
        \For{$j$ in $1:(i-1)$}{

           $\bar p\gets \bar p+A(i,j)/(n(n-1)/2)$\;

           $M \gets M+ A(i,j)(c(i)+c(j)-c(i)c(j))$\;
        
        }

}
}

\State{$T\gets \frac{M-\{\frac12k(k-1)+k(n-k)\}\bar p\bar \Delta}{\frac12n(n-1)\{\bar p(1-\bar p)\bar\Delta(1-\bar\Delta)\}^{1/2}}$}

\State \Return {$T, M$}
\EndFunction
\end{algorithmic}
\end{algorithm}

\begin{algorithm} 
\caption{Objective function update (adjacency matrix)}
\label{alg:update}
\begin{algorithmic}
\Require{$A$ adjacency matrix, $c$ core-periphery vector, $n$, $k$, $m$, $M$, node index $i$} 
\Ensure{$T$ objective function value, $M$ number of core-core and core-periphery edges, $c'$ core-periphery labels}
\Statex
\Function{update}{$A, c, M,i$}
\State {$k' \gets k -c(i)$}
\State {$\bar\Delta' \gets (k'(k'-1)/2+k'(n-k'))/(n(n-1)/2)$}
\State {$\bar p \gets m/(n(n-1)/2)$}
\State{$M'\gets M$; $c'\gets c$; $c'_i=1-c_i$}

\State{\For{$j$ in $1:(i-1)$}{
    $M' \gets M'+ A(i,j)(c'(i)-c(i))(1-c(j))$
}

}

\State{$T'\gets \frac{M'-\{\frac12k'(k'-1)+k'(n-k')\}\bar p\bar \Delta'}{\frac12n(n-1)\{\bar p(1-\bar p)\bar\Delta'(1-\bar\Delta')\}^{1/2}}$}

\State \Return {$T', M', c'$}
\EndFunction
\end{algorithmic}
\end{algorithm}

\begin{algorithm} 
\caption{Greedy algorithm (adjacency matrix)}
\label{alg:greedyA}
\begin{algorithmic}
\Require{$A$ adjacency matrix} 
\Ensure{$c$ core-periphery labels}
\Statex
\Function{greedyAdj}{$A$}
\State {Randomly initialize labels $c$}
\State{$T,M\gets${\sc objFunAdj}$(A,c)$}
\State { $run \gets 1$}
\State{\While{$run>0$}{
 Set $run\gets0$\; 
 
 Randomly order nodes\;
 
  \For{$i$ in $1:n$}{

  $T',M',c'\gets$\sc{update}$(A, c, M, i)$
  
  \If{$T' > T$}{
  
  $c\gets c'$; $T\gets T'$; $M\gets M'$\;

  $run \gets 1$\;
 }
   
  }
}
}

\State \Return {$c$}
\EndFunction
\end{algorithmic}
\end{algorithm}

\end{document}